\input{aipcheck}

\documentclass[
    ,final            % use final for the camera ready runs
  ]
  {aipproc}

\usepackage{amssymb}

\layoutstyle{8x11single}

\def\be{\begin{equation}}
\def\ee{\end{equation}}
\def\ba{\begin{eqnarray}}
\def\ea{\end{eqnarray}}

\def\sfrac#1#2{{\textstyle \frac{#1}{#2}}}

\begin{document}

\title{ Valence quark contributions for\\
the $\gamma N \to P_{11}(1440)$ transition}

%12.39.Ki 	Relativistic quark model
%13.40.Gp 	Electromagnetic form factors
%14.20.Gk 	Baryon resonances (S=C=B=0)


\author{G. Ramalho}{
  address={Centro de F{\'\i}sica Te\'orica de Part{\'\i}culas,
Av.\ Rovisco Pais, 1049-001 Lisboa, Portugal}
}

\author{K. Tsushima}{
  address={
Excited Baryon Analysis Center (EBAC) in Theory Center,
Thomas Jefferson National Accelerator Facility, Newport News,
Virginia 23606, USA\\
and\\
CSSM, School of Chemistry and Physics,
University of Adelaide, Adelaide SA 5005, Australia}
}

\begin{abstract}
A covariant spectator quark model is applied to
estimate the valence quark contributions
to the $F_1^\ast(Q^2)$ and $F_2^\ast(Q^2)$
transition form factors for the $\gamma N \to P_{11}(1440)$ reaction.
The Roper resonance, $P_{11}(1440)$, is assumed to be
the first radial excitation of the nucleon.
The model requires no extra parameters
except for those already fixed by the previous studies for the nucleon.
The results are consistent with the experimental data in the
high $Q^2$ region, and those from the lattice QCD.
Finally the model is also applied to
estimate the meson cloud contributions
from the CLAS and MAID analysis.
\end{abstract}

\maketitle

%%%%%%%%%%%%%%%%%%%%%%%%%%%%%%%%%%%%%%%%%%%%
%% MAINMATTER
%%%%%%%%%%%%%%%%%%%%%%%%%%%%%%%%%%%%%%%%%%%%

\section{Introduction}

Study of the meson-nucleon reactions is one of the 
most important research topics associated with modern
accelerators like CEBAF at Jefferson laboratory,
and defines new challenges for theoretical models.
Although the electroproduction of nucleon
resonances ($\gamma N \to N^\ast$) is expected
to be governed by the short range
interaction of quarks and gluons, i.e., perturbative QCD,
at low and intermediate four-momentum transfer squared
$Q^2$, one has to rely on some effective
and phenomenological approaches such as
constituent quark models.

The Roper [$P_{11}(1440)$] resonance is
particularly interesting among the many 
experimentally observed nucleon resonances.
Although usual quark models predict the Roper state
as the first radial excitation of the nucleon,
they usually result to give a much larger mass than
the experimentally observed mass~\cite{Aznauryan07}.
Experimentally the Roper has
a large width
which implies that it can be
a $\pi N$ or $\pi \pi N$ molecular-like system,
or alternatively, a system of a
confined three-quark core surrounded by
a large amount of meson clouds.
The system can be studied using
dynamical coupled channel models for
the meson-nucleon reactions~\cite{Burkert04}.

To describe the $\gamma N \to P_{11}(1440)$ transition, 
we use a covariant spectator model~\cite{Gross69,Nucleon,Omega,Octet,NDelta}.
The model has been successfully applied
to the nucleon~\cite{Nucleon,Octet,FixedAxis,ExclusiveR} and
$\Delta$ systems~\cite{NDelta,NDeltaD,LatticeD,DeltaFF}.
A particular interest of this work is the
model for the nucleon~\cite{Nucleon}.
In the covariant spectator quark model a baryon is described
as a three-valence quark system with an on-shell quark-pair or diquark
with mass $m_D$, while the remaining quark is off-shell and
free to interact with the electromagnetic fields.
The quark-diquark vertex is represented by  a baryon $B$ wave function
$\Psi_B$ that effectively describes quark confinement~\cite{Nucleon}.
To represent the nucleon system, 
we adopt the simplest structure given by a symmetric
and antisymmetric combination of the diquark states
combined to a relative S-state with
the remaining quark~\cite{Nucleon}:
\be
\Psi_N= \frac{1}{\sqrt{2}}
\left[ \Phi_I^0 \Phi_S^0 + \Phi_I^1 \Phi_S^1
\right] \psi_N(P,k)
\label{eqPsiN}
\ee
where $\Phi_{S}^{0,1}$ [$\Phi_{I}^{0,1}$] is the
spin [isospin] state which 
corresponds to the diquark with the quantum number 0 or 1.
The function $\psi_N$ is a scalar wave function
which depends exclusively on $(P-k)^2$,
where $P$ ($k$) is the baryon (diquark) momentum.
As the Roper shares the same spin and isospin quantum numbers
with the nucleon
its wave function $\Psi_R$
can also be represented by Eq.~(\ref{eqPsiN})
except for the scalar wave function $\psi_R$.

The constituent quark electromagnetic current in the model is described by
\be
j_I^\mu =
\left( \sfrac{1}{6}f_{1+} + \sfrac{1}{2}f_{1-} \tau_3 \right)
\left(\gamma^\mu - \frac{\not\!q q^\mu}{q^2}\right)+
  \left( \sfrac{1}{6}f_{2+} + \sfrac{1}{2}f_{2-} \tau_3 \right)
j_2 \frac{i \sigma^{\mu \nu}q_\nu}{2M},
\ee
where $M$ is the nucleon mass
and $\tau_3$ is the isospin operator.
To parameterize the electromagnetic structure
of the constituent quark in terms of the
quark form factors $f_{1\pm}$ and $f_{2\pm}$,
we adopt a vector meson dominance based parametrization
given by two vector meson poles: a light one
corresponds to the $\rho$ or $\omega$ vector meson,
and the heavier one corresponds to
an effective heavy meson with mass $M_h= 2M$ 
to take accounts of the short range phenomenology.
This parametrization allows us to extend 
the model for other regimes such
as lattice QCD~\cite{Omega,LatticeD,Lattice}.

The $\gamma N \to P_{11}$ transition in the model
is described by a relativistic impulse approximation
in terms of the initial $P_-$ and final $P_+$ baryon momenta,
with the diquark (spectator) on-mass-shell~\cite{Roper}:
\ba
J^\mu &=&  3 \sum_{\Lambda} \int_k \bar \Psi_R (P_-,k)j_I^\mu \Psi_N(P_-,k), 
\nonumber \\
&=& \bar u_R (P_+) \left[
\left(\gamma^\mu -\frac{\not \! q q^\mu}{q^2} \right)F_1^\ast(Q^2)
+ \frac{i\sigma^{\mu \nu} q_\nu}{M_R + M} F_2^\ast (Q^2)
\right]u(P_-).
\ea
In the first line
the sum is over the diquark states $\Lambda=\{s,\lambda\}$,
where $s$ and $\lambda=0,\pm1$ stand for
the scalar diquark 
and the vector diquark polarization, respectively,
and the covariant integral is $\int_k \equiv \int d^3k/[2E_D(2\pi)^3]$,
where $E_D$ is the diquark energy.
The factor 3 is due to the flavor symmetry.
In the second line the transition
form factors $F_1^\ast$ and  $F_2^\ast$ are defined
independent of the frame
using the Dirac spinors of the Roper ($u_R$) and nucleon ($u$)
with the respective masses $M_R$ and $M$.
For simplicity the spin projection indices
are suppressed.

%%%%%%%%%%%%%%%%%%%%%%%%%%%%%%%%%%%%%%%%%%%%
%% Sample figure:
%%
%% The option [height=...] scales the picture to the given height,
%% without it it would be printed at its nominal size
%%%%%%%%%%%%%%%%%%%%%%%%%%%%%%%%%%%%%%%%%%%%

\begin{figure}
% original size: [width=3.2in]
  \includegraphics[width=3.0in]{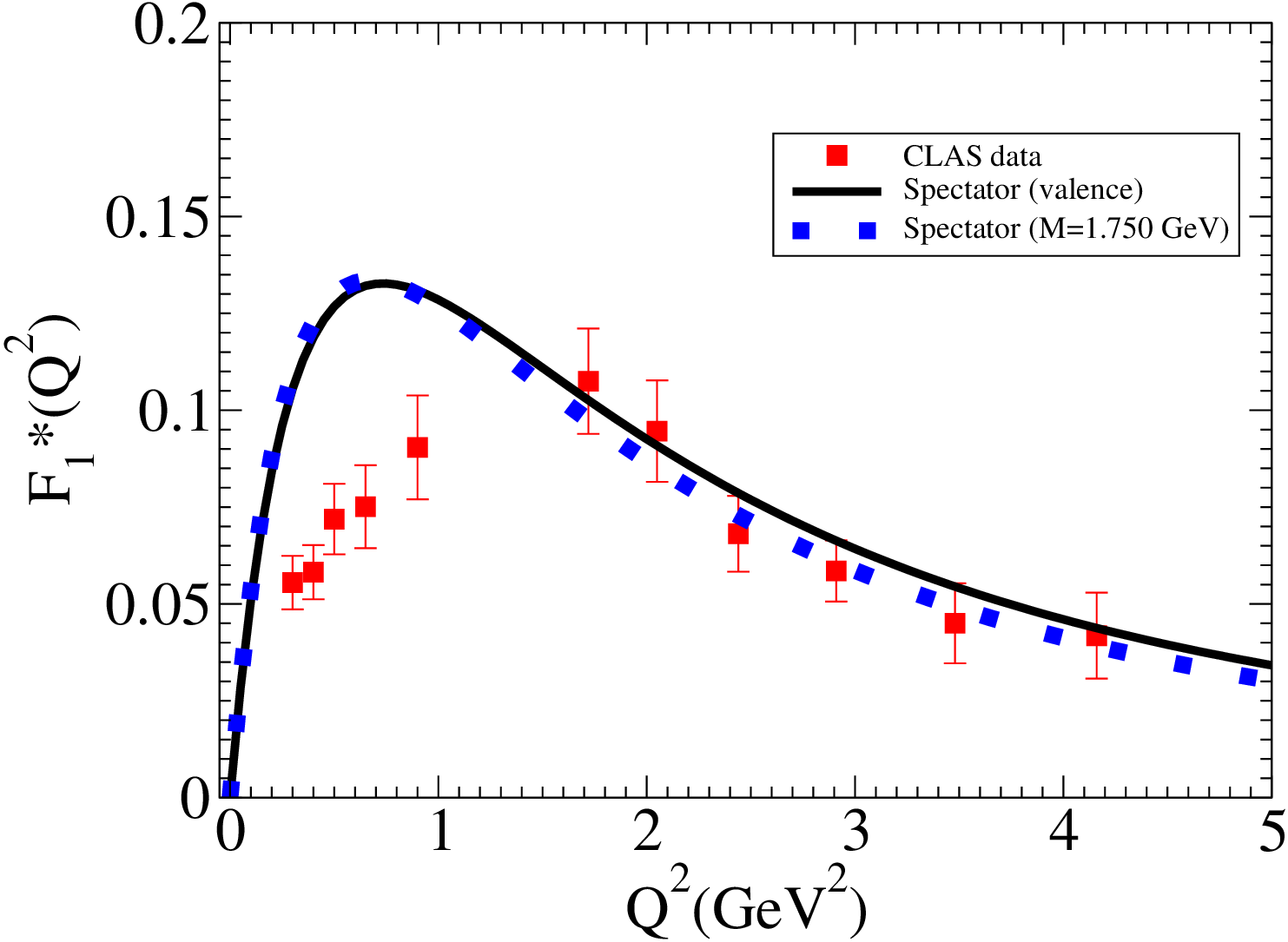}  \hspace{.1cm}
\includegraphics[width=3.0in]{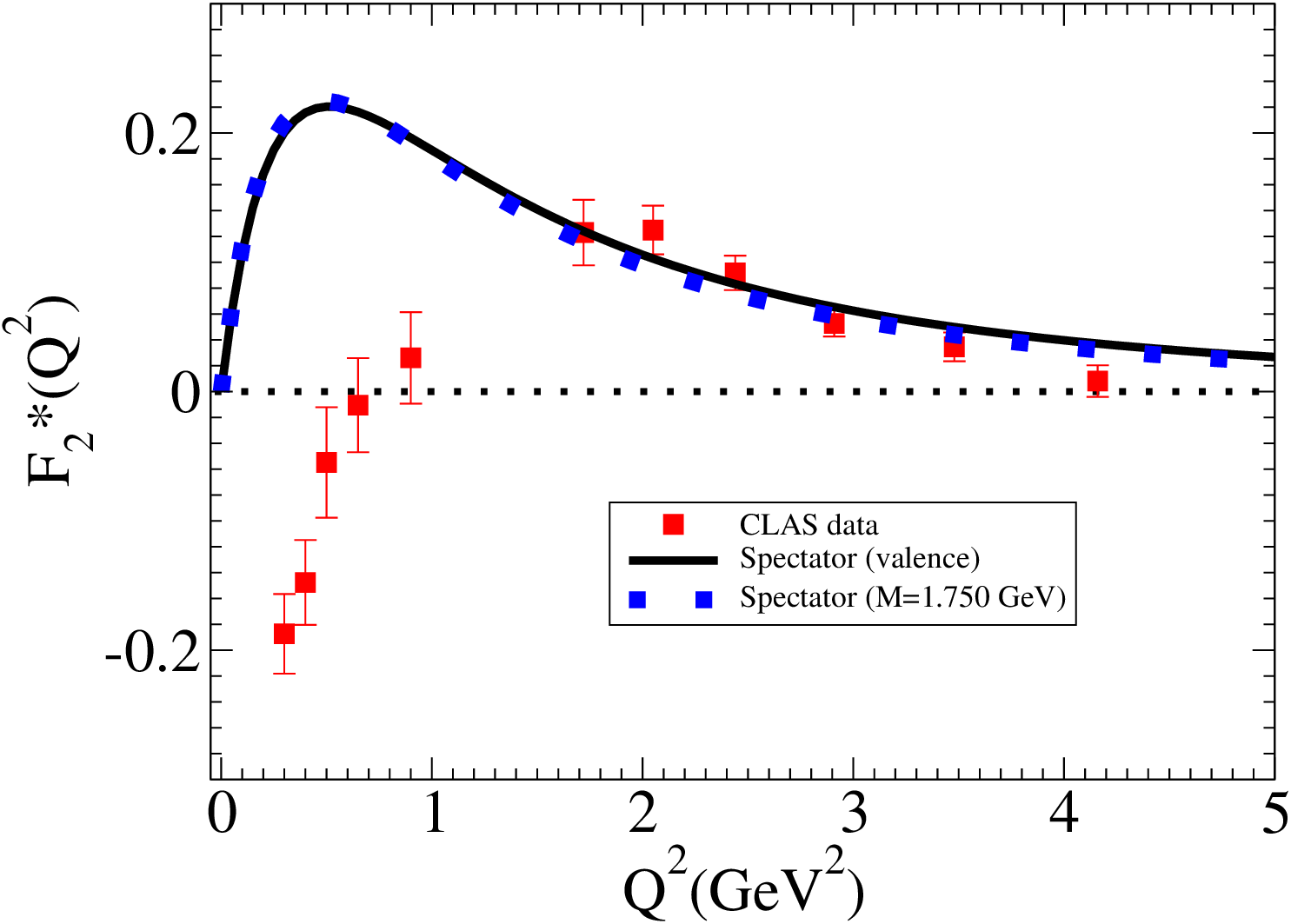}
  \caption{$\gamma N \to P_{11}(1440)$ transition form factors~\cite{Roper}.
The solid [dotted] line represents the result for $M_R=1.440$ GeV [1.750 GeV].
Data are taken from Ref.~\cite{CLAS}.}
\label{Roper}
\end{figure}

To represent the scalar wave functions as functions of $(P-k)^2$, 
we can use a dimensionless variable
\be
\chi_{_B}= \frac{(M_B-m_D)^2-(P-k)^2}{M_Bm_D},
\ee
where $M_B$ is the baryon mass.
With this form, the nucleon and the Roper scalar wave functions
are given by~\cite{Roper}:
\ba
\psi_N(P,k)&=& \frac{N_0}{m_D(\beta_1 + \chi_{_N})(\beta_2+\chi_{_N})}, \\
\psi_R(P,k)&=& \frac{\beta_3-\chi_{_R}}{\beta_1 + \chi_{_R}}
\frac{N_1}{m_D(\beta_1 + \chi_{_R})(\beta_2+\chi_{_R})}.
\ea
The nucleon scalar wave function
is chosen to reproduce the asymptotic form predicted
by pQCD for the nucleon form factors ($G_E,G_E \sim 1/Q^4$)
and to describe the elastic nucleon form factor data~\cite{Nucleon}.
The parameters $\beta_1$ and $\beta_2$
are associated with the momentum range.
With $\beta_2 > \beta_1$,
$\beta_1$ and $\beta_2$ are the long-range
and short-range regulators, respectively.
The expression for the Roper scalar wave function is inspired
by the nonrelativistic wave functions  
in the spherical harmonic potentials~\cite{Aznauryan07,Capstick95,Diaz04},
where the factor $\beta_3- \chi_R$
(linear in the momentum variable $\chi_R$)
characterizes the radial excitation.
The factors $N_0$ and $N_1$ are normalization
constants given by the condition $\int_k |\psi_B|^2=1$
at $Q^2=0$ for $B=N,R$.
The new parameter $\beta_3$, associated with the Roper,
is fixed by the orthogonality with the nucleon wave function.

\section{Results}

The results for the $\gamma N \to P_{11}(1440)$
transition form factors are presented in Fig.~\ref{Roper}.
Recall that all parameters associated with the Roper
have been determined by the relation with the nucleon.
Therefore, the results are genuine predictions.
The agreement with the data
is excellent for $Q^2>2$ GeV$^2$,
in particular for $F_1^\ast$.
Our results are also consistent with other work
for the intermediate and high $Q^2$ region.
See Ref.~\cite{Roper} for more details.
In dynamical coupled channel models~\cite{Suzuki10}
the Roper appears as a heavy bare system
($\approx$ 1.750 MeV) dressed by meson clouds that reduce
the mass of the system to near the observed mass
($\approx$ 1.440 MeV).
To explore further
this reduction of mass due to the meson clouds, 
we replace the
physical mass by the 'bare' mass.
The result is indicated by the dotted line.
As one can observe from the figure the effect
due to the Roper bare mass is small 
although it approaches very slightly to the data.

The disagreement in the low $Q^2$
region in Fig.~\ref{Roper}
can be interpreted as a limitation of the
approach, since only valence quark effects have been included, 
but not the quark-antiquark, or meson cloud effects.
The meson cloud effects are expected to be important
in the low $Q^2$ region~\cite{Aznauryan07,Burkert04,CLAS}.
This argument is supported by the fact that
when the model extended to a heavy pion mass lattice QCD regime,
where the meson cloud effect is suppressed, 
since the result agrees well the heavy pion mass 
lattice data~\cite{Roper,Lin09}.
The importance of the meson cloud contributions
in inelastic reactions was also observed
in the $\gamma N \to \Delta$ reaction~\cite{NDelta,NDeltaD,LatticeD}.
Furthermore, it was shown that the
covariant spectator quark model
can describe both the lattice and physical data~\cite{LatticeD}.
Based on the successes in describing
the heavy pion mass lattice QCD data
and the high $Q^2$ region data,
we have some confidence that the model can describe
well the valence quark contributions.
Thus, we can use the spectator quark model to estimate
the meson cloud contributions.

\begin{figure}[t]
%$\begin{array}{cc}
%\includegraphics[width=3.2in]{F1subMx}  & \hspace{.1cm}
%\includegraphics[width=3.2in]{A12e} \\
%\includegraphics[width=3.2in]{F2subMx}  & \hspace{.1cm}
%\includegraphics[width=3.2in]{S12d}
%\end{array}$
%\vspace{-.5c}
$\begin{array}{cc}
\includegraphics[width=3.0in]{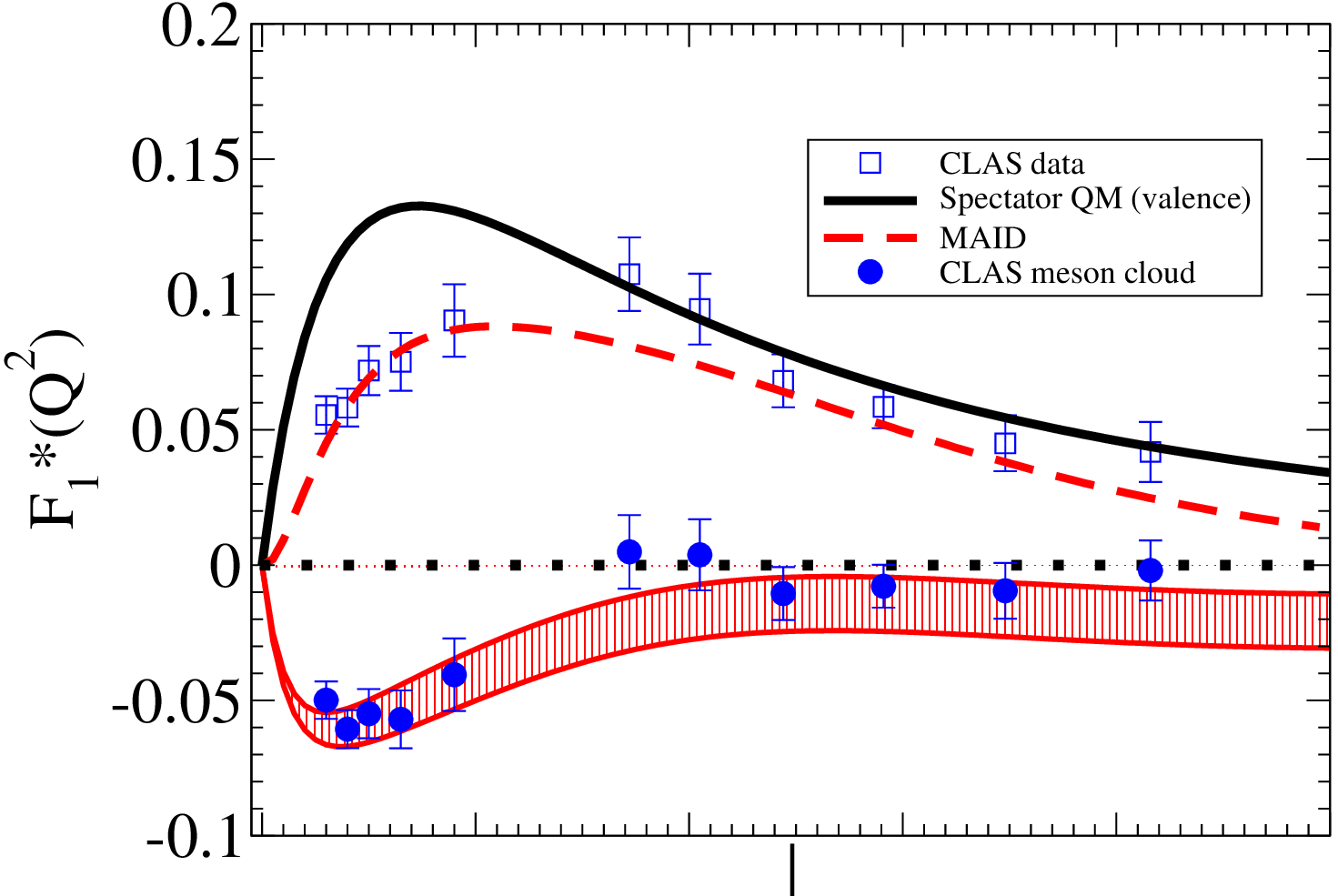} \vspace{.05cm}
& \hspace{.1cm}
\includegraphics[width=3.0in]{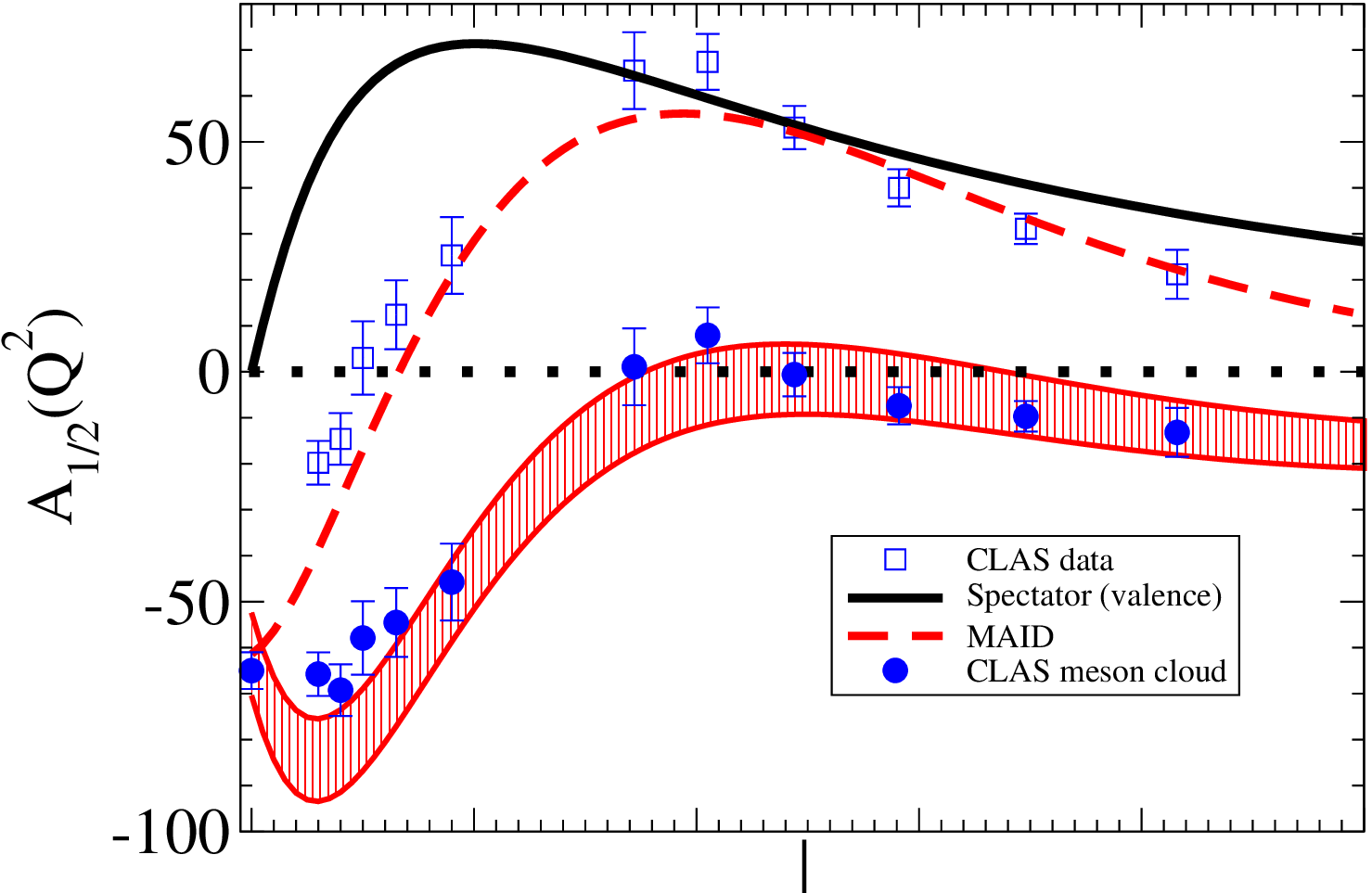}\vspace{.06cm} \\
& \\
\includegraphics[width=3.0in]{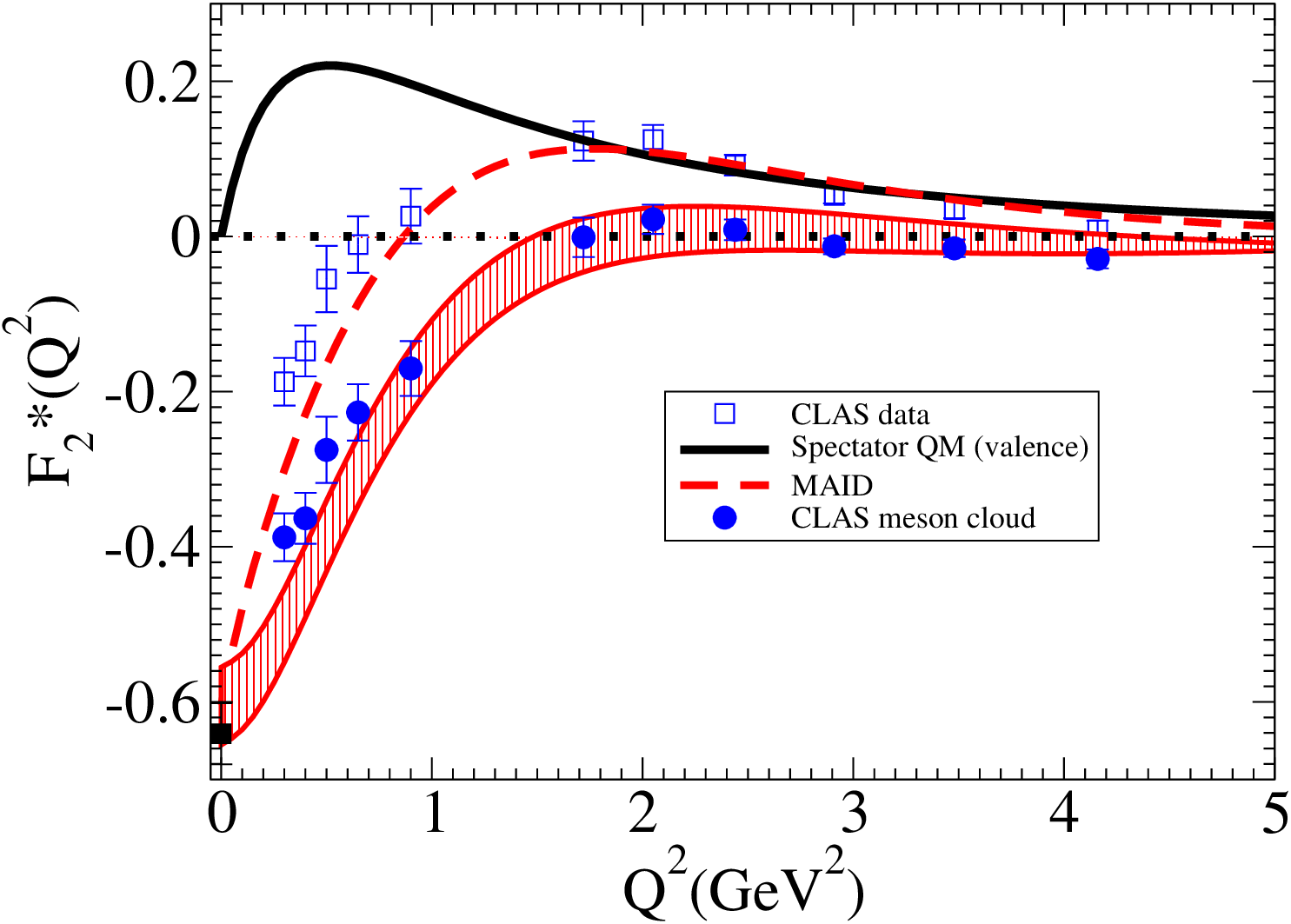}  & \hspace{.1cm}
\includegraphics[width=3.0in]{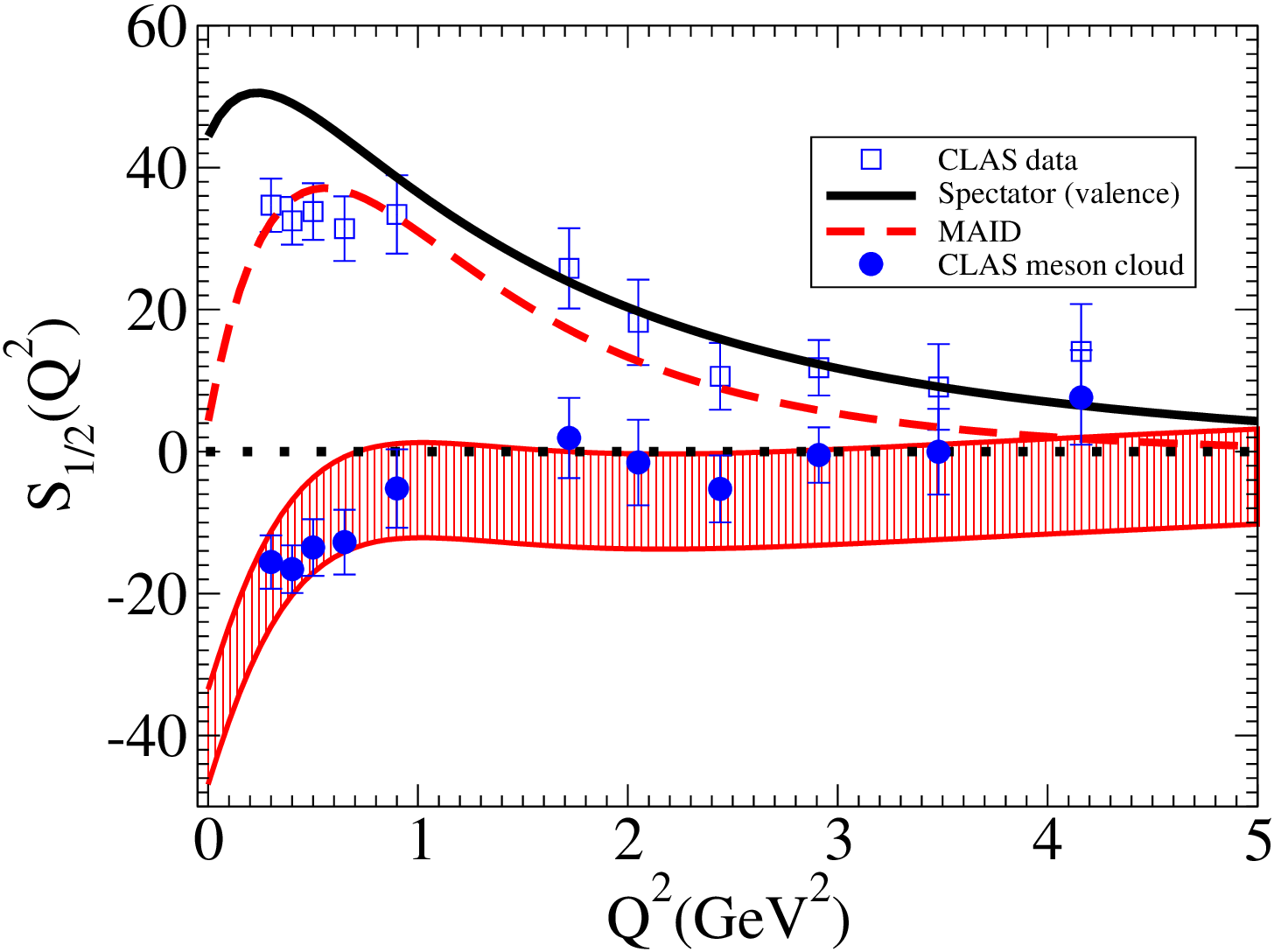}
\end{array}$
%\vspace{-.5cm}
  \caption{
$\gamma N \to P_{11}(1440)$ transition form factors (left) and
amplitudes (right) \cite{Roper},
compared with the CLAS data (squares)~\cite{CLAS},
and the meson cloud contributions estimated from CLAS data (circles).
The dashed line is the MAID fit from Ref.~\cite{Tiator09}.
The bands represent the limits of the meson cloud
contributions estimated from the MAID fit~\cite{Roper}.}
\label{Amp}
\end{figure}

To estimate the meson cloud contributions,
we decompose the form factors as,
\be
F_i^\ast(Q^2)= F_i^b (Q^2) + F_i^{mc}(Q^2) \hspace{.2cm} (i=1,2),
\ee
where $F_i^b$ and $F_i^{mc}$ represent
the valence quark (bare) and meson cloud contributions, respectively.
This decomposition is justified if the
meson is created by the overall baryon
but not by a single quark in the baryon core.
Replacing $F_i^b$ by the result of the spectator quark model, one can
estimate the meson cloud contributions in
the $\gamma N \to P_{11}(1440)$ transition.
We estimate the meson cloud contributions in two different ways as 
presented in Fig.~\ref{Amp} (left panel).
One is based on the MAID fit made for the whole $Q^2$ region,
and estimates the bands associated with the meson cloud assuming
the uncertainty of the CLAS data points (red region).
The other estimate is from CLAS data
for each data point by subtracting the
valence quark contribution. The results
are represented by circles.

Similarly, the meson cloud contributions
can be determined for the
helicity amplitudes, $A_{1/2}$ and $S_{1/2}$, associated
with the photon polarizations of $+1$ and $0$ respectively.
Representing the corresponding amplitudes by
$R=A_{1/2}$ and $S_{1/2}$, respectively,
one can obtain the meson cloud contributions:
\be
R^{mc}(Q^2)= R(Q^2) - R^{b}(Q^2).
\ee
The results are also shown in Fig.~\ref{Amp} (right panel).
Both methods suggest the significant
meson cloud effects for low $Q^2$ region ($Q^2 < 1$ GeV$^2$),
and a fast falloff with increasing $Q^2$.

In conclusion, the spectator quark model
can be effectively applied to study 
the nucleon resonances, particularly in the intermediate
and high $Q^2$ region, where the valence quark effects
are dominant, and both the relativity and covariance
are essential.
Another example of a successful application of the
present approach can be found in Ref.~\cite{Delta1600}.

%%%%%%%%%%%%%%%%%%%%%%%%%%%%%%%%%%%%%%%%%%%%%%%%
%% BACKMATTER
%%%%%%%%%%%%%%%%%%%%%%%%%%%%%%%%%%%%%%%%%%%%%%%%

\begin{theacknowledgments}
The authors would like to thank V.~D. Burkert for
the invitation.
G.~R. thanks the organizers for the financial support.
G.~R.\ was supported by the Portuguese Funda\c{c}\~ao para
a Ci\^encia e Tecnologia (FCT) under the grant
SFRH/BPD/26886/2006.
Notice: Authored by Jefferson Science Associates, LLC under U.S. DOE Contract No. DE-AC05-06OR23177. The U.S. Government retains a non-exclusive, paid-up, irrevocable, world-wide license to publish or reproduce this manuscript for U.S. Government purposes.
\end{theacknowledgments}



\bibliographystyle{aipproc}   % if natbib is available

%\bibliography{sample}

%\vspace{-.2cm}

%\vspace{-.2cm}

\newpage

\end{document}